# Improvements to Time Bracketed Authentication

by Charles H. Bennett
IBM Research, Yorktown Heights, NY 10598 USA
18 Aug 2003

We describe a collection of techniques whereby audiovisual or other recordings of significant events can be made in a way that hinders falsification, pre-dating, or post-dating by interested parties, even by the makers and operators of the recording equipment. A central feature of these techniques is the interplay between private information, which by its nature is untrustworthy and susceptible to suppression or manipulation by interested parties, and public information, which is too widely known to be manipulated by anyone.   While authenticated recordings may be infeasible to falsify, they can be abused in other ways, such as being used for blackmail or harassment; but susceptibility to these abuses can be reduced by encryption and secret sharing.

**I. The importance of video evidence and the problem of faking**.

Audiovisual documentation is becoming increasingly important in settling disputes. Every day it helps convict the guilty, exonerate the innocent, and reveal the subtleties of real world situations that interested parties may try to hide in biased, oversimplified verbal accounts of what happened. Perhaps more importantly, the mere knowledge that an A/V recording is, or *might be,* being made is a powerful disincentive to crime and abuse of power on all scales, depriving pickpockets and war criminals alike of the confidence that their misdeeds will go unrecorded and therefore unpunished. The most famous case of this in the US was the amateur video of Rodney King's beating by Los Angeles police in 1991, which led to an increased public awareness of police brutality and eventually to a conviction of some of the officers. Recent examples can be seen in the BBC news story [1] about amateur videos of arrests in Oklahoma and California.  Even though videos have been used to expose police misconduct, many police departments have embraced the mounting of video cameras in police cars and the routine videotaping of arrests, interrogations and confessions, in order to produce more convincing evidence and protect officers from false accusations of misconduct.

Unfortunately, the same digital technology which makes today's video cameras small and convenient seriously undermines the reliability of their recordings. Given enough time, a skilled digital counterfeiter can add or remove persons or objects from a scene, alter the words a person speaks and the actions they perform, and in the end produce a realistic looking video that bears no resemblance to the truth. For example, Chris Bregler et al's Video Rewrite [2] shows how to make videos of people saying things they never said, like John F. Kennedy saying "I did not inhale." The usual way to prevent such falsification has been to record the video under controlled circumstances and thereafter keep it continuously under guard until it is used as evidence. Unfortunately, amateur videos do not have this protection, and professional videos are often made and guarded by parties, e.g. police, who are far from disinterested, and who therefore may be suspected of having falsified the video themselves. For example, in the O.J. Simpson murder trial, the police were accused of falsifying video and other evidence. A recent approach, obviating the need for guarded storage, is the use of a watermarking camera [3] that digitally signs and time stamps images as they are made. But this can be defeated by collusion with the camera manufacturer, or by physically tampering with the camera after manufacture.

## II. Time-bracketed authentication

Time bracketed authentication [4] provides a way of making videos that remain trustworthy even if the manufacturer and operators of the video equipment are untrustworthy. It works by combining and improving two old ideas:

- The kidnapper's trick of photographing a hostage holding today's newspaper to prove the hostage is still alive. We call this a *past time bracket*;

- The inventor's practice of publishing an invention, or putting it in a self-addressed envelope and getting it postmarked, to prove the invention originated no later than a certain day. We call this a *future time bracket*.

Both ideas both involve communication with an *outside world* beyond the control of the distrusted party: in the first case, information is imported from a place beyond the kidnapper's control; in the second case information is exported to a place beyond the inventor's control.

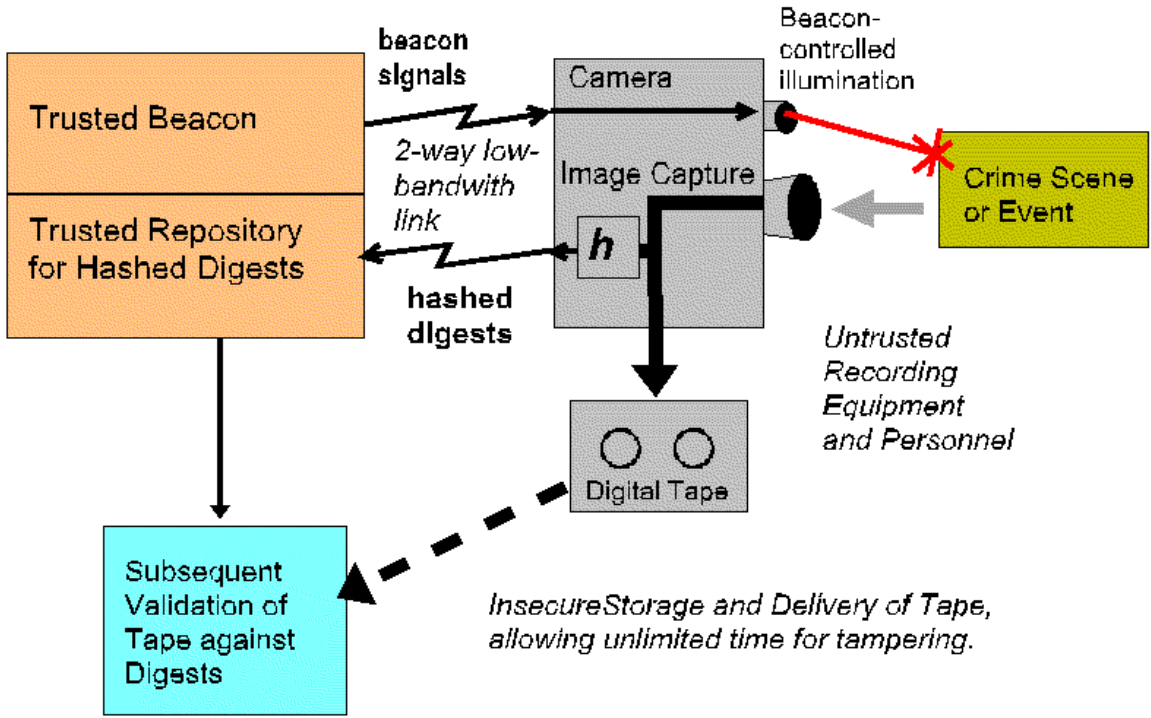

# Time Bracketed Authentication

In more detail (cf Fig. 1) time-bracketed authentication works by importing a stream of unpredictable information from an trusted public source, or **beacon** [5] and using it to influence the scene being recorded, e.g. by having an LCD projector or laser near the camera illuminate the scene with a beacon-controlled texture or scan pattern. At frequent intervals, the most recent portion of the A/V data stream, compressed or uncompressed, which is about to be appended to the cumulative record on tape or other output medium, is compressed by the secure hash function $h$, and the resulting hashed digest exported to trusted **repository**, i.e. a public archive or other place beyond the control of any interested party. This having been done, the videotape, or other digital record, need not be guarded or protected from subsequent manipulation, because if it were modified outside the short time interval between arrival of the beacon signal (which provides a past time bracket) and archiving of the first digest influenced by it (which provides a future time bracket), it would no longer match the archived digests. The only remaining opportunity for falsification is for the would-be falsifier to use a fast, real-time rendering program capable of simulating fake action, including the effect of the unpredictable beacon-controlled illumination, within the short time interval mentioned above. If the scene being videotaped includes a cooperative human subject or interviewee, the beacon signals can be caused to influence the scene in more complex ways, for example by directing the subject to perform, on camera, a action from a menu of possible actions, the timing and choice of the action being controlled by the beacon signal. The actions on the menu should be easy to perform, but hard to render (e.g. "touch your nose", "clap your hands"...) and might employ hard-to-render props ("blow some soap bubbles").

Given a trusted beacon and trusted repository, the grounds for believing the past and future time brackets are rather different. The plausibility of the future time bracket follows simply from the presumed infeasibility of finding collisions under the hash function $h$. The past time bracket is more subtle, depending also on the presumed infeasibility of simulating (rendering) the beacon's influence in real time. If it were possible to do this simulation, the video recording, while matching all its archived digests, might still be a fake, made in a computer by pure animation or by manipulating prerecorded material in the manner of Video Rewrite. Past time bracketing thus involves not only mathematics and computational complexity but also physics and perhaps even biomechanics and psychology.

Future time bracketing by itself, proving that a document was not created or modified after a certain date, is already quite useful. Its theory was developed in classic papers of Haber et al on digital time stamping [6], and it is commercially available under the name "digital notary service".

The following section III explores the notions of beacon and repository and ways of implementing them in practice. Section IV discusses ways of using one data modality (e.g. audio) at the scene being recorded to influence another (e.g. video) so as to further hinder the falsification of either. Section V describes techniques for making authenticated recordings more *discreet,* that is resistant to being used for wrongful purposes such as harassment and blackmail, while retaining their value as evidence in legitimate circumstances. Section VI describes how amateur videographers, who unexpectedly find themselves in a position to document an important event like the Rodney King beating, can apply the principles of time-bracketed authentication in an impromptu fashion to considerably improve the evidentiary value of their recordings.

## III. Beacons, Repositories, and the Public/Private Interface

In legal and official settings there has long been a notion of public information, or matters of public record, for example birth and death certificates and transcripts of legislative deliberations.   Such information is expected to be stable over time and accessible to everyone.  Its resistance to loss or alteration comes from being kept in a well-guarded location, or distributed widely to many places, or both.   In recent years the Internet, cheap storage, and efficient search engines have made publication and access to information much cheaper and faster. Sites such as www.arXiv.org undertake to provide a free permanent repository of scientific preprints. The similarly named site www.archive.org maintains snapshots of the entire publicly-accessible portion of the Web dating back to 1996.  Newsgroup postings dating back to the early 1980s appear set to remain permanently accessible through archive sites such as Google.groups. Of course a website's content can be changed, and an arXiv or newsgroup submission can usually be retracted ("nuked") by its originator, but if the change is at all controversial, it engenders a trail of comment, which itself becomes part of the public record, recoverable through search engines.   A famous example is the Pathfinder news site's accidental posting, for a few seconds, of a page erroneously announcing a guilty verdict in O.J. Simpson's murder trial [7]. For an authenticated public record, of course, it is important that publication be permanent, with entries being neither retractible nor alterable by the originator or anyone else.  In this paper we consider a somewhat idealized world in which, for at least some identifiable types of information, a phenomenon of *irreversible publication* exists: what has once been published remains accessible ever afterward, and cannot be changed or deleted.  The invention of printing, and several centuries later, the proliferation of cheap digital storage media and the World Wide Web, have made this idealization closer to reality than it was in the ancient world, or even twenty years ago.

Although much public information is meaningful and indeed carefully crafted by its originators, an important complementary feature of public information is that some of it is unpredictable and beyond anyone's control. Low order digits in newspaper reports of financial or sports data, for example, are so unpredictable that bets are taken on them in the gambling game known as the numbers racket.   The distinctive feature of such information is that it is unknown and uncontrollable by anyone before a certain time, but shortly thereafter becomes universally accessible and unalterable. Sources of unpredictable public information have proliferated on the Web, including live webcasts of sports events and live webcams (e.g. http://lemenille.com/traffic_cam_gwb.html depicting traffic at the George Washington Bridge in Fort Lee, New Jersey at one minute intervals).

Irreversible publication, to the extent it exists today, is affected by social and legal factors, as well as technological ones. While it is economically feasible for a website to publish and archive many megabytes of data per day, either locally generated or collected from outside submitters, such a site could not offer automatic permanent publication of all content submitted to it without worrying about copyright and libel. Existing archives and newsgroups typically require submitters to promise not to post libelous, infringing, or other types of objectionable material and they sometimes remove postings upon receiving a credible complaint that the policy has been violated. However, if, for one reason or another, the data stream is known a priori to be non-objectionable, then the site could safely promise permanence. For example, a traffic cam website can be reasonably confident that the data it publishes is unobjectionable (except perhaps to commuters during rush hour). Either the original operator of a traffic site, or anyone else on the Web, could at little expense archive the traffic cam's output, and make it available for later retrieval, when it might have evidentiary value, for example in the investigation of a traffic accident happening within view of the camera.

If the aim of an archive is merely authentication, rather than publication of content, then a secure hash function can be used to guarantee that the postings, while useful for authentication, will not be objectionable. Hash functions have long been used [8] to attest the integrity digital documents and files and are central to the digital time-stamping techniques of Haber et al [6], where they are used to construct a chain or tree of hash values connecting any particular document to a master hash value placed into the public record, for example as a paid newspaper ad. The chain itself would not necessarily be published, but would rather be given to the client, who could then authenticate document at will, without further cooperation of the time stamping service, by producing the document and the chain of hash values connecting it to the public record. We will use hash functions in a slightly different way, taking into account the greater ease of publication in the Internet era. Let a hash-and-publish (HAP) service be an agency or mechanism (e.g. an automated website) that takes client submissions *s,* applies a secure one-way hash function *h,* and publishes the hash value

*v=h(s),* along with the time *t* it was submitted. The random-seeming string *v* would have only a negligible chance of being objectionable, but would serve to authenticate the fact that *s* had been submitted, when and if, at some later time, the submitter wished to prove that fact. The HAP would thus provide, not publication or nor even, strictly speaking, authentication, but rather an *option to authenticate*, exercisable at will by the submitter. By using anonymous proxy techniques, submitters could access a HAP anonymously, without disclosing their identity. Submitters, especially anonymous ones, would generally not want to send meaningful content to a HAP; instead they would first hash their content *c* to obtain a hash value *s=h(v)*, which they would send to the HAP, who would then hash it again and publish the pair *(v,t)=(h(s),t)=(h(h(c)),t).* The first hashing serves to protect the submitter from unwanted disclosure of meaningful content by a malicious or incompetent HAP; the second hashing, as described before, protects the HAP from inadvertently publishing objectionable material submitted by a malicious or incompetent submitter.

A simple HAP-based distributed implementation of a trustworthy time-stamping service would have the user, who wishes to authenticate a string *s* at time *t*, send it (anonymously, if desired) to each of several geographically and administratively separate HAPs. Each HAP would each publish, e.g. by posting on a website publicly accessible forever afterward, the pair *(t,v)*, where *v=h(s)*. The posting and continued availability of the pair on the majority of the HAPs (that is, all but the dishonest or defective ones) would constitute evidence that *s* existed at time *t* and had not been modified afterward.

As noted earlier time bracketed authentication needs not only a trusted repository, such as a set of independent HAPs, but a trusted source of public randomness, here called a random beacon. In time bracketed authentication the beacon helps to frustrate postdating, but beacons have other uses, for example in synchronizing transactions and choosing lottery numbers. The examples of the numbers racket and traffic cams have already been mentioned. It is technically fairly easy to extract true random digits, approximating arbitrarily closely a sequence of ideal independent coin tosses, from probabilistic physical processes such as radioactive decay or noisy electrical circuits, and several such true random number generators, or TRGs, are commercially available. Using one of these (or a home-made TRG if one distrusts the manufacturers) it would not be overly expensive to set up a beacon site that would generate and publish up to several thousand random bits per second, while maintaining a web-accessible archive of its previously emitted digits and their times of emission. To deny itself the temptation to lie about past emissions, the site could allow itself to be mirrored elsewhere on the Web, have its output stream regularly time stamped [6], and/or publish the data in an offline nonvolatile medium such as an annual CD. The main weakness of a beacon is not that it might lie about its past outputs, but rather that it is impossible in principle to convince outsiders, who did not participate in building and operating the beacon site, that the numbers it publishes are truly unpredictable, rather than being known long beforehand by a dishonest beacon operator and accomplices, who might take advantage of their insider information to defeat the cryptographic purposes of the beacon. For a beacon to be useful for purposes such as lotteries or time bracketed authentication, not only the user, but the general public, must be confident of its fairness.

A good way to protect against dishonest or defective beacons is similar to the technique recommended above for protection against defective HAPs—namely, one should use a number of geographically and administratively separate beacons. If these beacons are well synchronized and spacelike separated (so they cannot influence one another) it suffices to take their XOR to obtain a bit stream that will be random and unpredictable if at least one of the contributing beacons is honest and properly functioning. In [5] we described how, with beacons that are not spacelike separated, and therefore potentially capable of influencing one another, one can still obtain a sequence that is at least partly unpredictable by timesharing among the beacons rather than XORing them. A better solution is to have each beacon pre-commit, at each discrete time *t*, to the string it plans to emit at some future time *t*+$\delta$, where $\delta$ is a fixed delay. In more detail, such a precommitting random beacon (PCRB) at each discrete time *t*, generates a new random bit string *r(t)* of some suitable length (say 128 or 256 bits) using its TRG, and stores this string secretly and locally for later emission at time *t*+$\delta$. Meanwhile, at time *t*, the beacon pre-commits to *r(t)* by emitting *h(r(t))*, where, as usual, *h* denotes a one-way collision-free hash function. The full emission at time *t* is thus the triple (*t, h(r(t)), r(t-$\delta$)*). A typical Web-based implementation of a PCRB might post such triples every second of coordinated universal time where $\delta$ could be of the order of a few seconds. The PCRB website would also be expected to keep its past emissions available online, allow itself to be mirrored or time stamped, and publish an offline backup in the form of an annual CD.

To obtain beacon data more trustworthy than that coming from any one beacon, a user would, at each time, XOR together the emissions $\{r_i : i=1\ldots m\}$ of several independent PCRBs of the sort just described, meanwhile checking for each one that its currently emitted random string $r_i(t\text{-}\delta)$ agrees with the previously emitted commitment $h(r_i(t\text{-}\delta))$, and excluding from the XOR any beacon failing this test.   The resultant string $r_1 \oplus r_2 \oplus r_3\ldots$  will then be trustworthy if at least one of the PCRBs on which it is based is honest and correctly functioning.

PCRBs and HAPs are alike in that they are services functioning at the public/private interface whose reliability can be increased by having users combine multiple geographically and administratively separate suppliers of the service.  Nonperformance by a HAP would be easy to detect, if not prevent, and would consist either of its failing to hash and publish in a timely manner, or in its attempting to lie about its past emissions.   The latter fault could be neutralized by wide initial publication of the output stream, approximating the ideal of irreversible publication.  Nonperformance by a PCRB would include these detectable faults, but also the undetectable fault of producing data that is not unpredictable, and has been leaked secretly beforehand to accomplices.

Since PCRBs' and HAPs' trustworthiness depends on multiplicity and administrative independence, they should be constructed using open source software and, in the case of a PCRB, open-source plans for building the hardware true random number generator (TRG) from widely available simple components such as discrete resistors, capacitors, and diodes.   This minimizes the chance that would-be operators of honest PCRBs will inadvertently build dishonest ones, for example by buying clandestinely sabotaged ready-made TRGs from a dishonest manufacturer.  To raise the small amounts of money needed to operate their websites, while maintaining administrative independence from any central authority, operators of PCRBs and HAPs might solicit voluntary donations or sell advertising to be displayed to website visitors and users.

A well constructed time-bracketed authentication (TBA) system would consist of a video camera or other recording equipment communicating in real time with several of independent PCRBs and HAPs.  The incoming signals from the PCRBs would be XORed together as described above to generate the trusted random influences (challenges) called for by the TBA protocol. Meanwhile the stream of hash digests generated by the TBA protocol would be sent to a multiple HAPs, where they would be hashed again and published.

**IV. Coupling Data Modalities to Hinder Falsification**

Some data modalities (e.g. audio) are easier to falsify than others (e.g. video). This can seriously impair the evidentiary value of audiovisual recordings where, for example, the sound track of a videotaped police interview might be susceptible to replacement (or allegations of replacement) by another sound track with entirely different meaning. This can be prevented to some extent by showing close-ups of the speakers' faces and mouths during spoken testimony, in effect using lip-reading to validate the correctness of the soundtrack. But this approach is difficult to implement if there are multiple persons speaking out of turn, and even if a clear view of the subject's mouth is available, it is often possible to find a visually similar phrases with quite different sound and meaning. When popular American TV shows are adapted for broadcast in Germany, audio track editors have been quite successful in finding German dialog which approximates the meaning of the English, and which, when spoken by trained German voice artists, appears to be in good synch with the original video track of American actors speaking English.

To obtain greater resistance to falsification, even in situations where the speakers' faces are not in view, we propose linking the audio with the harder-to-falsify video, by using light sources whose intensity, color, or position is modulated by the audio signal to provide some, but not all, of the light illuminating the scene. In police interviews and other similar adversarial situations, the visible modulation of the illumination in response to the speech of all the participants would give each participant some assurance that the audio signal being recorded corresponds to what they are actually saying. This technique, which we call audio modulated illumination (AMI) could be used in a standalone fashion, or in conjunction with time-bracketed authentication to further hinder falsification of the audiovisual recording.

When used in a standalone fashion, audio-modulated illumination would yield a video track containing visible evidence of the soundtrack, mixed in a complicated and hard-to-render fashion with non-audio-modulated illumination. For example, in one experiment with this technique, the speaker's voice varied the brightness of an incandescent lamp in the foreground, casting time-dependent highlights and shadows, while other parts of the scene, illuminated by light from an open window, remained at constant brightness.

When used in conjunction with time bracketed authentication, the audio signal could be used to control some features of the illumination, while the external random challenges from one or more beacons would be used to control others. A very simple way to do this would be to turn on an ordinary radio receiver during a recording session illuminated by AMI. The illumination would then respond both to the speakers' voices and the outside radio signal, which, if the broadcast were live rather than prerecorded, would serve as a beacon. This would combine the advantages of time-bracketed authentication (reducing the time available for falsification) with those of audio-modulated illumination (giving the participants confidence, even during the recording session itself, that the audio was being recorded correctly and not being secretly replaced by other audio).

More generally the technique of using a more easily falsifiable data modality to influence a less easily falsifiable one can be used in any situation where one wishes to make a trustworthy multimodal recording, with or without TBA, but where one of the data modalities is dangerously easy to falsify on its own. A typical example would be an industrial process, where time-varying sensor data such as temperatures in different parts of the plant, (which of course could be falsified relatively easily if they were recorded solely in a separate data track) would also be used to modulate the illumination of the video track in a complex and hard-to-falsify manner.

## V. Enhancing discretion of authenticated recordings against improper use or disclosure

While people generally recognize the value of surveillance videos for deterring and prosecuting wrongdoing, there is also great concern that pervasive surveillance, or even the proliferation of amateur videographers, could lead to harassment and blackmail of people for things they get recorded doing that are not wrong or illegal, but merely embarrassing, or properly regarded as nobody's business but their own. This becomes all the more true if the video or other recording is made in a less than public venue, where one or more participants has a legitimate expectation of privacy. The issue is the subject of established legislation and precedent in the case of recordings of phone conversations, and the question of privacy of email is also an active topic of discussion. Broadly speaking one can say that in any setting, there will be an expectation, which may be codified legally, of some level of privacy and access rights depending both on the place and on the persons present. For example, in a bank or casino one expects one's actions and speech to be watched and recorded. In one's own home, conversely, one expects there will be no surveillance without a court order. Likewise, in a two-way telephone conversation the expectation is that barring a court-ordered wiretap the conversation will be private unless both parties consent to the recording.

To continue with the example of a telephone conversation, it may happen that, after the fact, both parties wish that a recording had been made, for example to help resolve a dispute over what they said, or to prove to a third party that they said some particular thing. The ability to generate such a recording after the fact, but only if both parties agree, can be approximated cryptographically, by setting up the telephone equipment to automatically make an encrypted recording of all conversations, using a separate random encryption key for each conversation, and dividing the encryption key into two shares (using the well known technique of secret sharing), one of which would be given to each participant in the conversation. The encrypted recording would be given to both participants, but it could only be decrypted if both parties consented, which is the same as the condition presently prevailing for making an unencrypted recording in the first place.

Using this kind of cryptographic protection it would be possible, and not technically difficult, to implement a more discreet form of authenticated surveillance, whereby encrypted (and if desired TBA-authenticated) recordings would be routinely made in various situations (e.g. phone conversations, airport lobbies, street corners, diplomatic negotiating sessions), in each case with an access structure, enforced by secret sharing, corresponding to people's codified expectations of privacy. When used with TBA as in Figure 1, the digital tape at the center right would have been automatically encrypted by the equipment producing it, so unless the equipment had been tampered with, no unencrypted version of the data would remain. The same equipment would generate the requisite shares of the encryption key and distribute them to all parties having rights to participate in the decision of whether to decrypt the tape. The parties involved in this secret sharing need not be static. Using short range radio transponders similar to those used in automatic toll plazas, one could even set up an authenticated conversation room where the access structure would change dynamically according to who was in the room: all conversations would be recorded, but no portion of the recording could be opened except by consent of all people in the room at the time.  The same dynamic access control could be implemented even more easily in a multi-way telephone conversation.  In any of these situations, the access structure could be set up to allow the usual quorum to be overruled by a court of sufficiently high authority.

## VI. TBA-Lite, a version of Time Bracketed Authentication suitable for amateur videographers and other impromptu users

It is anticipated that time bracketed authentication, with the various improvements described above, will one day be routinely used by professionals in settings requiring authentication, such as police interviews, courtroom testimony, surgical procedures, and formal negotiating sessions. Such professional users will presumably use a commercially supplied system consisting of hardware and software designed to automatically link with available PCRBs and HAPs,  apply the beacon-supplied signals to influence the a/v recording according to the TBA protocol, thereby producing hard-to-fake recordings under the particular conditions of use.   Another category of professional users, including photojournalists and police on patrol, would use a more versatile kind of professional TBA system designed to adapt at a moment's notice to a variety of operating conditions, indoor and outdoor, with different ambient light and sound levels.

However, a third important arena for application of these ideas would be in informal settings, by amateur videographers and others who have no special equipment beyond a standard consumer camcorder, digital or film still camera, or audio tape recorder (plus other widely available equipment such as a TV or radio receiver and a personal computer with standard business/personal software), but who unexpectedly find themselves at the right time and place to document something important.   These amateur users, who function essentially as impromptu photojournalists, can still apply the principles of time bracketed authentication, and the result will be a video or other recording which, while not of professional quality, will still provide much greater evidentiary reliability than if it had simply been recorded and put in a drawer.  To do take advantage of TBA principles, it suffices for the amateur videographer to use available equipment in a way that creates a past and future time bracket sufficiently close together in time that it is unlikely that the recording could have been significantly manipulated in the intervening time. As noted earlier, the past time bracket involves information flowing in from the outside world, beyond the control of the videographer, and influencing the video in a way that would be hard to fake quickly. The future time bracket involves exporting the video recording itself, or a digital hash of it, to the outside world beyond the control of the videographer.

For sufficiently large newsworthy events, such as earthquakes and train wrecks, the event provides its own past time bracket: the event was unexpected and unpredictable at the time it happened, but soon afterward news of it, including hard-to-fake details such as the collapse of particular buildings, pass into the public domain beyond the control of the videographer or other personnel at the scene. In other less momentous situations, for example an interview with a person not publicly known to have been at a particular place at a particular time, a deliberately generated past time bracket is advisable to establish that the recording was not made at some earlier time. An easy way to do this is to use a radio or TV receiver tuned to a live (and therefore unpredictable beforehand) broadcast of news, weather, traffic conditions, or sports, any of which can serve as an improvised beacon. To avoid the problem of sound tracks being relatively easy to substitute without changing the video, the TV receiver could be turned so as to illuminate, or create reflections in, objects in the scene being recorded. Alternatively, if there is a human subject, he/she could be asked to react visibly in real time to the content of the TV/radio broadcast, for example applauding whenever his/her favorite sports team scores a point.

To establish a future time bracket it is sufficient to quickly place the videotape or other recording (eg audiotape, still photograph), or preferably several copies of it, into the public arena (eg by mailing it to the offices of one or more independent news media), where it can no longer be modified by the videographer or other interested parties. This approach can be used both with modern digital recordings and with the formerly widely used analog formats such as film cameras, still or video, and analog electronic audiotape or videotape. It has the disadvantage of requiring that a somewhat bulky and perhaps incriminating or irreplaceable material object, such as a videotape cassette or a roll of exposed film, be shipped quickly out of the battle zone into a safe place, away from the interested parties. If, on the other hand, the video or other recording is made in, or subsequently converted to, a digital format, and if the videographer has access to a personal computer, it suffices for the videographer to send a hash of it into the public arena. This can be done through almost any means of communication, even those available to civilians under wartime conditions. For example, if neither mail service nor telephone nor Internet were available, an amateur videographer, having clandestinely recorded evidence of a war crime in progress, could bury the original videotape in a secret location for later retrieval in safer times. But to establish a future time bracket, he/she could take a 64 bit (16 hex digit) hash of the video file, print a thousand copies of it on little slips of paper, and scatter these anonymously around the neighborhood or disperse them in the wind. Subsequent testimony by many people that the slips had been found on a certain date would establish that the videotape, although it might not be dug up until much later, must have been made before the date of discovery of the slips. Or the hash code, perhaps further truncated or converted to a more easily memorable mnemonic phrase or alphanumeric character string, could be propagated as an anonymous word-of-mouth rumor, with the admonition that it ought to be remembered and spread until it reached the outside world, because it preserves the evidentiary value of a buried videotape. The people spreading the rumor would not need to know sensitive information like the identity of the videographer or the location of the buried tape.

Under typical conditions the amateur techniques we have described, which we call TBA-Lite, would serve to bracket the time with a matter of minutes or hours. Professional time bracketed authentication, with connection via the Internet to PCRBs and HAPs and dedicated hardware and software implementing the TBA protocol, would of course create a much narrower time bracket on the order of a few seconds, or even less if the connection to the PCRBs and HAPs were made via radio rather than through the Internet. However, even a one-day time bracket considerably reduces the scope of plausible falsification, beyond what it would be if no effort were made to apply time-bracketing principles.

**Acknowledgments** I thank John Smolin, David DiVincenzo, Ralph Linsker, Daniel Gottesman, Ashish Thapliyal, and Manuel Blum for stimulating discussions of many of these ideas.

 **References**

[1] British Broadcasting Company BBC News video available online at http://news.bbc.co.uk/media/video/38128000/rm/_38128791_beatings07_cheallaigh_vi.ram (2002)

[2] C. Bregler, M. Covell, and M. Slaney "Video rewrite: Driving visual speech with audio" SIGGRAPH '97 Proceedings, Los Angeles CA (1997); Video rewrite website http://graphics.stanford.edu/~bregler/videorewrite/

[3] US Air Force site http://www.afmc.wpafb.af.mil/HQ-AFMC/PA/news/archive/2001/aug/Rome_Watermarkingcamera.htm describes a prototype watermarking camera from Eastman Kodak (2001).

[4] C.H. Bennett, D.P. DiVincenzo, and R. Linsker "Digital recording system with time-bracketed authentication by on-line challenges and method for authenticating recordings" US Patent 5764769 (1998), also discussed in reference [5], figure 1.

[5] C.H. Bennett and J.A. Smolin "Trust Enhancement by Multiple Random Beacons" eprint http://www.arXiv.org/ps/cs.CR/0201003 (2002).

[6] Stuart Haber and W. Scott Stornetta, "How to Time-Stamp a Digital Document". Journal of Cryptology, 3(2):99--111, 1991; Dave Bayer, Stuart Haber and W. Scott Stornetta "Improving the Efficiency and Reliability of Digital Time-Stamping"; Commercial digital notary service offered by http://www.surety.com .

[7] British website recounting erroneous "Guilty" news Web postings immediately after the OJ Simpson verdict http://www.darkin.demon.co.uk/html/ojguilty.html

[8] D.W. Davies and W.L. Price, "Security for Computer Networks". John Wiley & Sons, New York, 1984 (ISBN 0 471-90063 X), pp 136-137 describes authentication without a secret key by means of a one-way hash function.